\documentclass[12pt]{article}
\usepackage[T2A]{fontenc}
\usepackage[cp1251]{inputenc}
\usepackage[english]{babel}
\usepackage[dvips]{graphicx}
\normalsize
\usepackage{amssymb,amsmath}

\sloppypar
\newcommand{\teff}{$T_{\rm eff}$}
\newcommand{\eexc}{$E_{\rm exc}$}
\newcommand{\nnlte}{$N_{\rm i,non-LTE}$}
\newcommand{\nlte}{$N_{\rm i,LTE}$}
\newcommand{\dnlte}{$\Delta_{ \rm non-LTE}$}
\newcommand{\abnlte}{log~A$_{\rm non-LTE}$}
\newcommand{\ablte}{log~A$_{\rm LTE}$}

\def\vt{$\xi_{\rm t}$}

\newcommand{\kH }{$S_{\rm H}$}    

\begin{document}
\selectlanguage{english} 
\title{Influence of Collisions with Hydrogen on Titanium Abundance 	Determinations in Cool Stars}

\author{T. ~M. ~Sitnova \thanks{sitnova@inasan.ru}\;\;$^{1,2}$, S. A. Yakovleva$^2$, A. K. Belyaev$^2$, L. ~I.~Mashonkina$^1$ \\[2mm]
\begin{tabular}{l}
 $^1$ {\it \small{Institute of Astronomy, Russian Academy of Sciences, ul. Pyatnitskaya 48, Moscow, 119017 Russia}}\\[2mm]
 $^2$ {\it  \small{Herzen State Pedagogical University, Kazanskaya ul. 5, St. Petersburg, 191186 Russia}}
\end{tabular}
}
\date{}
\maketitle

\begin{abstract}
We performed the non-local thermodynamic equilibrium (non-LTE) calculations for Ti~I-II with the updated model atom that includes quantum-mechanical rate coefficients for inelastic collisions with hydrogen atoms.
We have calculated for the first time the rate coefficients
for bound-bound transitions in inelastic collisions of titanium atoms and ions with hydrogen atoms and
for the charge-exchange processes: Ti~I + H $\leftrightarrow$ Ti~II + H$^-$ and Ti~II + H $\leftrightarrow$ Ti~III + H$^-$.
The influence of these data on non-LTE abundance determinations has been tested for the Sun and four
metal-poor stars. For Ti~I and Ti~II, the application of the derived rate coefficients has led to an increase in
the departures from LTE and an increase in the titanium abundance compared to that, obtained with
approximate formulas for the rate coefficients. In metal-poor stars, we have failed to achieve consistent 
non-LTE abundances from lines of two ionization stages.
The known in the literature discrepancy  in the non-LTE abundances from Ti~I and Ti~II lines in metal-poor stars
cannot be solved  by improvement of the rates of inelastic processes in collisions with hydrogen atoms in non-LTE
calculations with classical model atmospheres.

\noindent
{\bf DOI:\/}  10.1134/S1063773720010041


\end{abstract}
\maketitle
\twocolumn
\section{Introduction}

In stars over a wide range of spectral types, from K
to A, titanium is observed in spectral lines of two ionization stages (Ti~I and Ti~II). Titanium lines can be
used to determine stellar atmosphere parameters --
the effective temperature (\teff) and the surface gravity
(log g). 
For accurate abundance determinations  from lines of Ti~I and Ti~II, the departures from local thermodynamic equilibrium (non-LTE effects) should be taken into account (Bergemann 2011, Sitnova et al. 2016).
In stellar atmospheres with \teff\ > 4500 K, neutral titanium is a minority species, and its statistical equilibrium (SE) can easily deviate from thermodynamic equilibrium owing to deviations of the mean intensity of ionizing radiation from the Planck function. 
 Overionization
of low-excitation levels by ultraviolet (UV) radiation, which leads to an underpopulation of atomic
levels and a weakening of spectral lines compared
to the LTE, is responsible for the departures from LTE for Ti~I. 
The accuracy of the non-LTE calculations depends on the completeness and quality of the
employed atomic data. In non-LTE calculations for Ti~I-II (Bergemann 2011; Bergemann et al. 2012; Sitnova
et al. 2016; Sitnova 2016; Mashonkina et al. 2017),
inelastic collisions with hydrogen were taken into account using the Drawin formula (Drawin 1968,
1969; Steenbock and Holweger 1984) with  a scaling coefficient to the rates (\kH). 
The application of this formula is criticized in the literature, since it does not contain the essential
physics and significantly overestimates the collision rate (Barklem et al. 2011).
However, it is used by astrophysicists
for non-LTE calculations in the absence of accurate
quantum-mechanical calculations.

Bergemann (2011) and Sitnova et al. (2016)
showed that non-LTE leads to agreement between
the abundances from Ti~I and Ti~II lines in stars
of spectral types from G to A and nearly solar
metallicity. However, for metal-poor halo stars, [Fe/H]\footnote{We use the standard notation for the elemental abundance ratios [X/H] = $\log(N_{\rm X}/N_{\rm tot})_{star} - \log(N_{\rm X}/N_{\rm tot})_{Sun}$.} < $-2$, with well-known atmospheric parameters the non-LTE abundances from
Ti~I and Ti~II lines have not been reconciled neither
with \kH\  = 1 (Sitnova et al. 2016) nor with \kH\  = 3
(Bergemann 2011). The  discrepancy
in the non-LTE abundances from Ti~I and Ti~II
lines as a function of metallicity was discussed by
Sitnova (2016). 
For dwarf stars in the solar neighborhood,
consistent abundances from two
ionization stages were obtained for stars with [Fe/H] > $-2$, while stars with a lower metallicity show a
discrepancy in the mean non-LTE abundances from
Ti~I and Ti~II lines, $\Delta_{\rm TiI-TiII}$, up to 0.35 dex. A similar behavior of
$\Delta_{\rm TiI-TiII}$ was  found for cool giants by Mashonkina et al. (2017). Due to the lower
effective temperatures of giants, the abundances from
two ionization stages begin to diverge at a lower
metallicity, [Fe/H] < $-3$.

At present, the non-LTE calculations  in metal-poor stars incorrectly predict the departures from LTE for Ti~I and, possibly,
also for Ti~II. The departures
from LTE increase with decreasing metallicity due
to a rise in the UV flux and a decrease in the rates
of collisions with electrons. Because of the lower
electron density, the accuracy of data for the
collisions with hydrogen atoms plays a crucial role
in non-LTE calculations for metal-poor stars. In
this paper, for the first time 
we present the rate coefficients of excitation and de-excitation processes for bound-bound transitions in collisions
 of titanium atoms and ions with hydrogen atoms and for the following charge-exchange
processes: Ti~I + H $\leftrightarrow$ Ti~II + H$^-$ and Ti~II + H $\leftrightarrow$ Ti~III + H$^-$. These results are of great practical importance for modeling the formation of titanium lines
in the atmospheres of cool stars.

Our calculations of the rate coefficients for transitions in inelastic collisions of titanium atoms and
ions with hydrogen atoms are presented in Section~\ref{rates_ti}.
A brief description of the stellar sample, employed for
testing the impact of the derived rate coefficients
on the non-LTE abundance is given in Section~\ref{obspar}.
The method of calculation of the spectra is described in
Section~\ref{method}. The results of our non-LTE calculations
and the derived non-LTE abundances are
presented in Sections  \ref{seti} and \ref{abundances}, respectively.

\section{Rate coefficients for inelastic processes in collisions with hydrogen}\label{rates_ti}

The rate coefficients were calculated for excitation and de-excitation processes
for bound-bound transitions in collisions of titanium
atoms and ions with hydrogen atoms and for the
following charge-exchange processes: Ti~I + H $\leftrightarrow$ Ti~II + H$^-$ and Ti~II + H $\leftrightarrow$ Ti~III + H$^-$. For our
calculations we used the quantum model approach
proposed by Belyaev and Yakovleva (2017a, 2017b)
and based on the application of the asymptotic semi-empirical
approach and the two-channel Landau-Zener model within the Born-Oppenheimer approach.
This approach allows us to calculate the rate
coefficients for the inelastic processes associated with
the transitions due to the long-range ionic-covalent
interaction of electronic states in the quasi-molecules
formed in collisions of atoms and ions of various
chemical elements with hydrogen atoms and anions.

When considering collisions of Ti~I with hydrogen,
we included 107 molecular states of the TiH
quasi-molecule in our calculations, two of which correspond
to the ionic pairs: Ti~II (3d$^2$4s $^4$F) + H$^-$ and Ti~II (3d$^3$ $^4$F) + H$^-$. The model approach allows
only the single-electron transitions between various
states of the same molecular symmetry
of the quasi-molecule to be taken into account;
therefore, we performed two sets calculations of the rate
coefficients and included the Ti~I (3d$^2$4s\,nl $^{3,5}$L) + H and Ti~II (3d$^2$4s $^4$F) + H$^-$
 states in one of them and
the Ti~I (3d$^3$nl $^{3,5}$L) + H and Ti~II (3d$^3$ $^4$F) + H$^-$
states in the other one. Note that the Ti~I (3d$^3$4s $^{3,5}$L)
states were included in both analyses. The transitions
between electronic states can occur within various
molecular symmetries. However, since both ionic
configurations produce molecular states of symmetries
 $^4\Sigma^-$, $^4\Pi$, $^4\Delta$, and $^4\Phi$, the covalent states of
only the same molecular symmetries were included
in our analysis. The transitions occurring within
each molecular symmetry were considered separately,
while the rate coefficients for each inelastic process
were summed over the molecular symmetries.

\begin{figure*}  
	(а)	\resizebox{70mm}{!}{\includegraphics[width=0.8\linewidth,trim=0.0cm 0.0cm 12.0cm 0cm,clip]{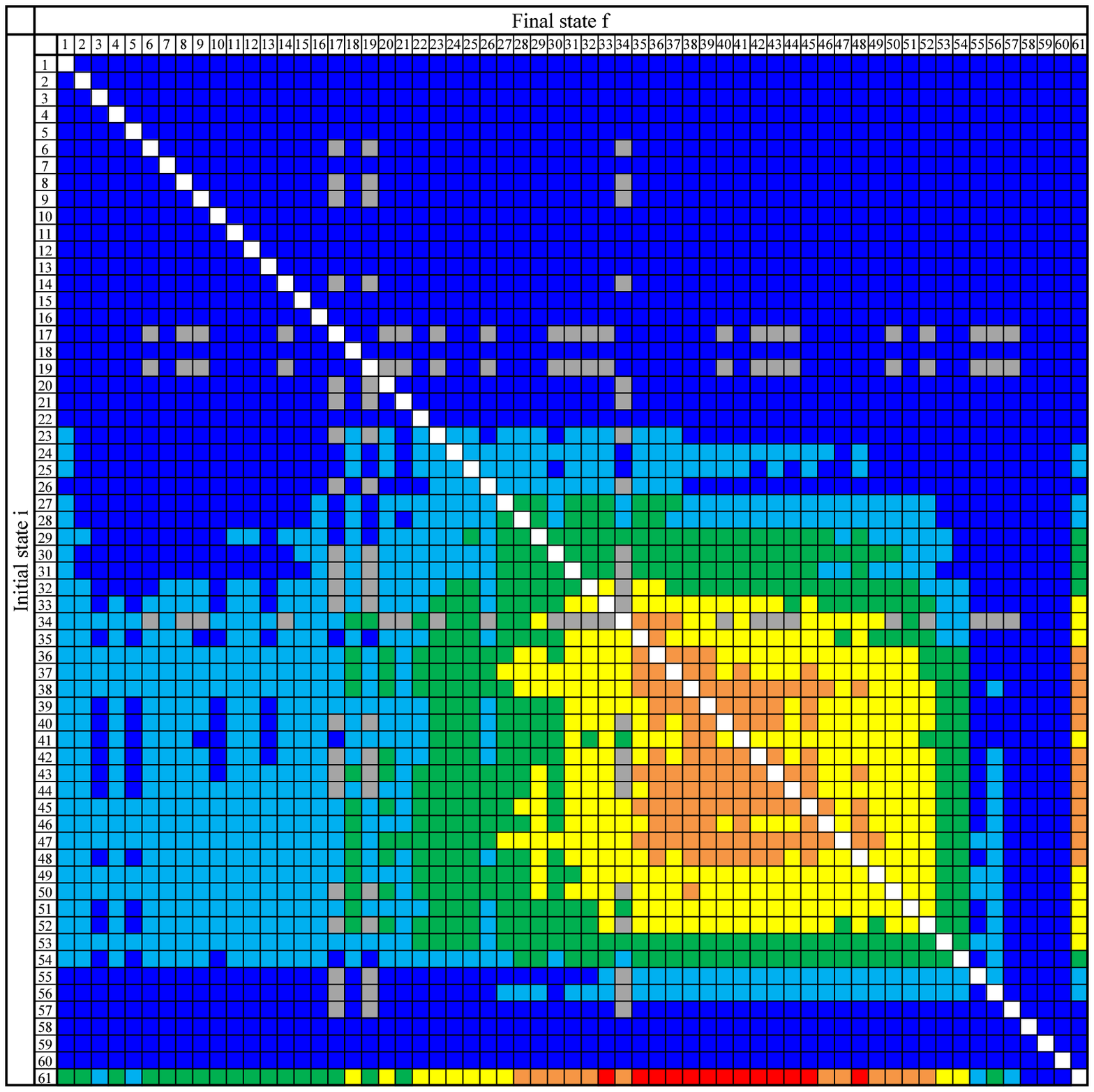}}
	(b)	\resizebox{70mm}{!}{\includegraphics[width=0.8\linewidth,trim=0.0cm 0.0cm 9.70cm 0cm,clip]{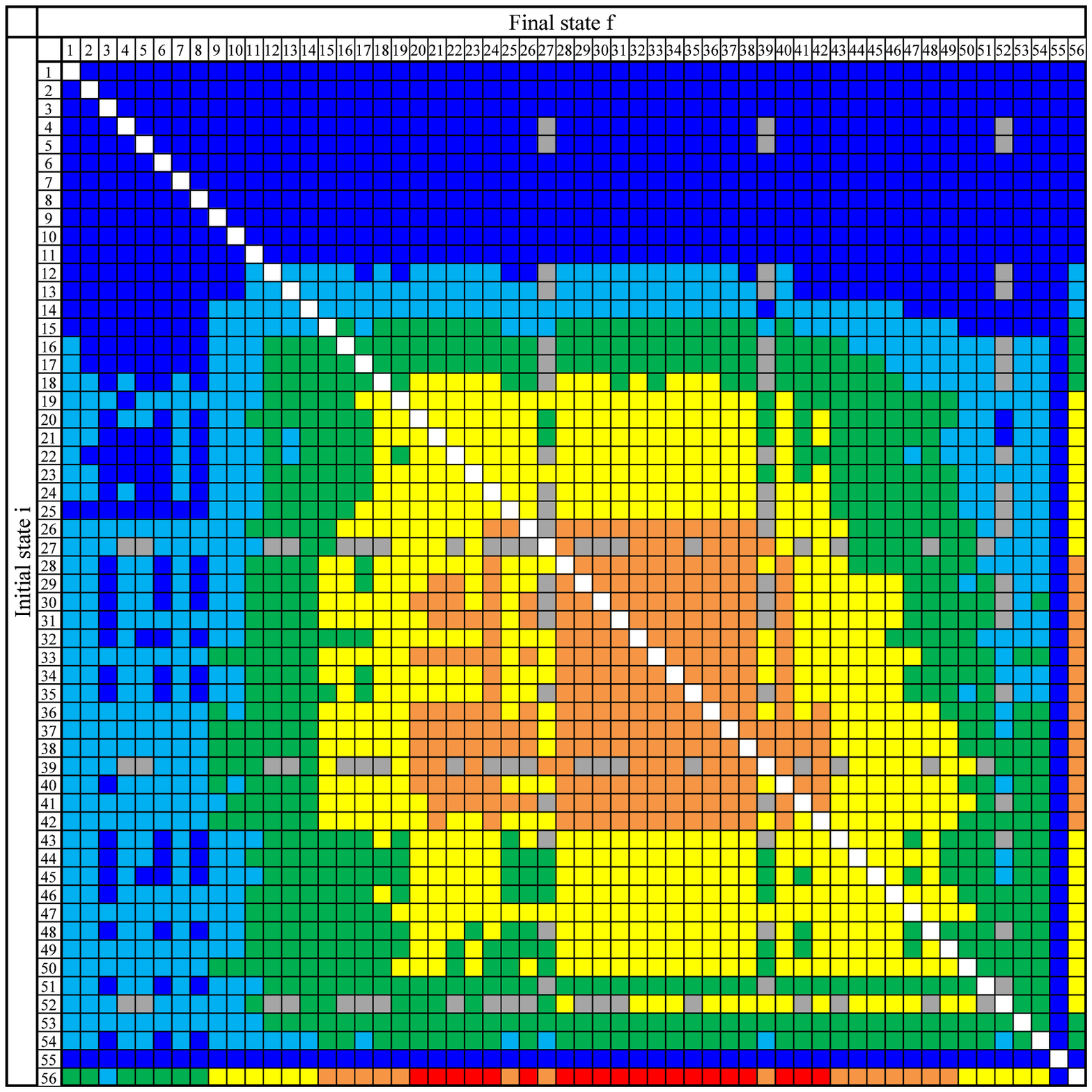}}	
	(c)	\resizebox{150mm}{!}{\includegraphics[trim=0.0cm 0.0cm 5.0cm 0cm,clip]{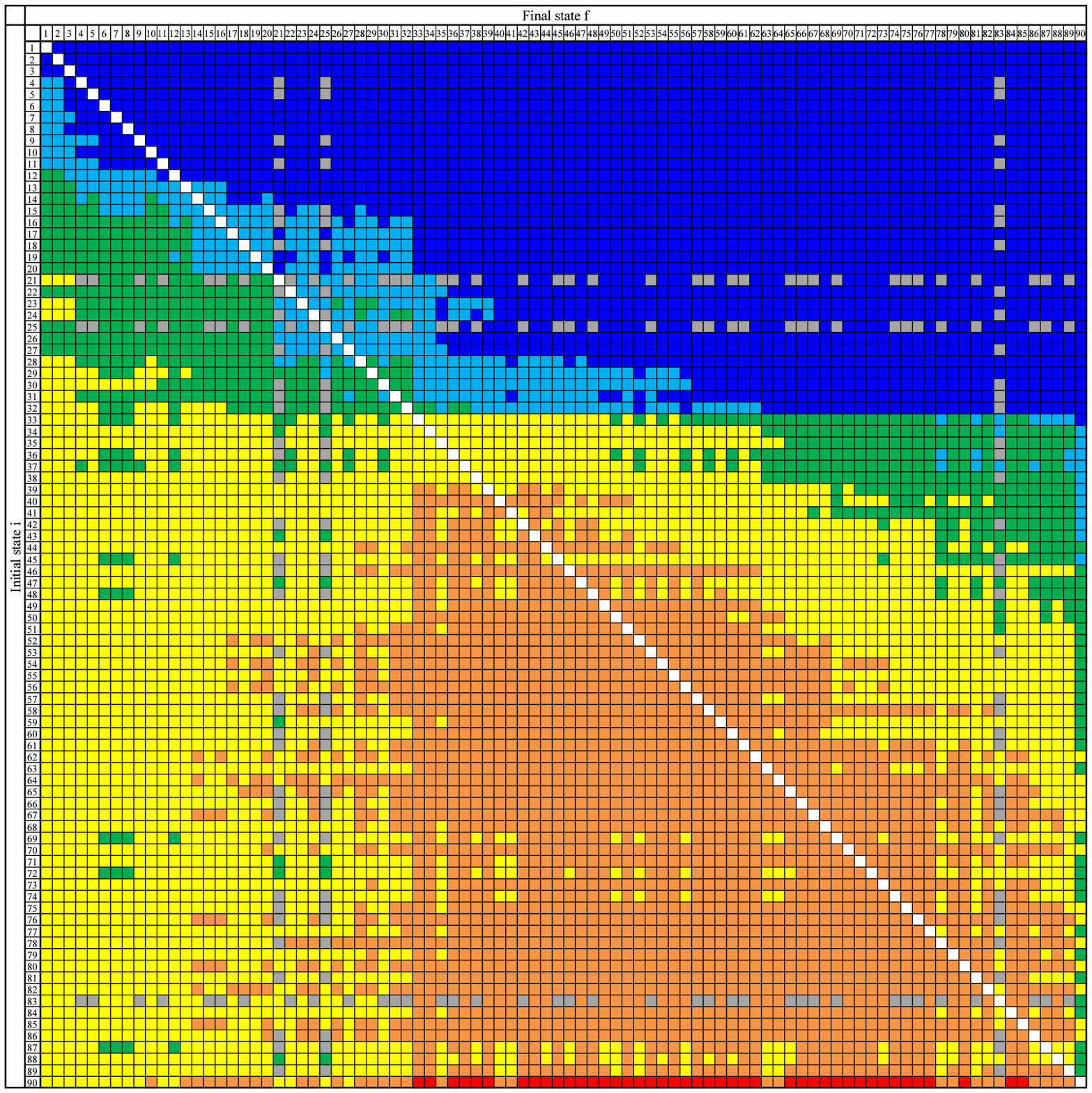}}	
	\caption{
		Graphical representation of the rate coefficients for the inelastic processes  in the collisions of
		Ti~I (a, b) and Ti~II (c) with hydrogen. The rate coefficients with values > 10$^{-8}$, 10$^{-9}$ -- 10$^{-8}$, 10$^{-10}$-- 10$^{-9}$, 10$^{-11}$-- 10$^{-10}$,
		10$^{-12}$-- 10$^{-11}$, and < 10$^{-12}$ cm$^3$ s$^{-1}$ are indicated by the red, orange, yellow, green, cyan, and blue colors, respectively.
		} 
	\label{fig:TiH_1}
\end{figure*}

When investigating collisions of Ti~II with
hydrogen, we considered 90 molecular states of the
TiH$^+$ quasi-molecule, including one state corresponding
to the ionic pair Ti~III (3d$^2$ $^3$F) + H$^-$. Since
the ionic state has molecular symmetries $^4\Sigma^-$, $^4\Pi$, $^4\Delta$, and $^4\Phi$, our calculations were performed
within each of the molecular symmetries.
Graphical representations of the calculated rate
coefficients for all inelastic processes are presented
in Fig.~\ref{fig:TiH_1}. The rate coefficient in this figure is
indicated by a color from red (with values greater
than 10$^{-8}$ cm$^3$/s) to blue (with values less than
10$^{-12}$ cm$^3$/s). Note that the rate coefficients for
elastic processes were not calculated in this study
and, therefore, they are indicated by the white color,
while the rate coefficients for the processes associated
with the transitions between states of different molecular
symmetries are zero and indicated by the gray
color.

The two upper panels show the rate coefficients
for the processes occurring in the collisions of Ti~I
with hydrogen: Figs. 1a and 1b present, respectively,
the rate coefficients for the processes associated with
the transitions due to the interaction of the covalent
Ti~I (3d$^2$4s\,nl $^{3,5}$L) + H  states with the first ionic
Ti~II (3d$^2$4s $^4$F) + H$^-$ state and due to the interaction
of the Ti~I (3d$^3$nl $^{3,5}$L) + H states with the
second ionic  Ti~II (3d$^3$ $^4$F) + H$^-$ state. We formed a
complete matrix of rate coefficients for the transitions
between all 107 states considered to be used in our
subsequent calculations. We performed summation
for the excitation and deexcitation processes associated
with the transitions between the Ti~I (3d$^3$4s $^{3,5}$L) states occurring due to the interaction with both
ionic configurations. The rate coefficients for the
processes associated with the transitions between the
states interacting with different ionic configurations
are zero. Figure 1c presents the rate coefficients for
the inelastic processes occurring in the collisions of
Ti~II with hydrogen.

It can be seen from the graphical representations
in Fig. 1 that the largest rate coefficients correspond
to the processes of mutual neutralization into the
states for which the electron binding energy
is in an optimal window and the processes associated
with the transitions between these states.
As shown by Belyaev and Yakovleva (2017a, 2017b),
the electron binding energy in such states is about
2 eV for the collisions of neutral atoms with hydrogen,
corresponding to Ti~I excitation energies in the range
4-5 eV, and about 4 eV for the collisions of singly
charged positive ions with hydrogen, corresponding
to Ti~II excitation energies in the range 8-10 eV.

The derived rate coefficients are accessible at
 http://www.non-lte.com/ti\_h.html and in the arXiv source file.

\section{Stellar sample, atmospheric parameters, and observations}\label{obspar}

The sample of stars includes four metal-poor
stars and the Sun. Their atmospheric parameters
are given in Table~\ref{param_tab}. For HD~84937, HD~140283,
and HD~122563 we use the atmospheric parameters
as in Mashonkina et al. (2019). Here we 
only note that based on photometric measurements
(Casagrande et al. 2010), stellar angular diameter
measurements (Karovicova et al. 2018; Creevey
et al. 2012, 2015), and the distances (Bailer-Jones
et al. 2018) derived from the trigonometric parallaxes
from the Gaia DR2 archive (Brown et al. 2018), the
effective temperature and surface gravity 
for these stars are reliably fixed within 50 K and
0.05 dex, respectively. For the ultra metal-poor
giant CD-38245, stellar parameters  were determined in Sitnova et al. (2019),
 using the color-magnitude calibrations, distances, isochrones, and the
non-LTE analysis of the Balmer line profiles and the
Ca I/Ca II ionization equilibrium.

For abundance determinations, we used high spectral 
resolution ($\lambda/\Delta\lambda$~>~45~000) observations
with a signal-to-noise ratio S/N > 50. The spectra
were taken from the UVES\footnote{http://archive.eso.org/eso/eso\_archive\_main.html} archive. For HD~84937 and HD~140283, we also used
the spectra in the UV range (1875 -- 3158 \AA), obtained at the Hubble Space Telescope
with the STIS spectrograph. The UV spectra were
reduced by T. Ayres and are publicly accessible
at http://casa.colorado.edu/$\simeq$ayres/ASTRAL. The
solar abundance was determined using a spectrum of
the Sun as a star (Kurucz et al. 1984).

\begin{table*}[htbp]
	\caption{Atmospheric parameters and titanium abundances for the sample stars}
	\label{param_tab}
	\tabcolsep1.0mm
	\begin{center}
		\begin{tabular}{crrrrrrrr}
			\hline\noalign{\smallskip}
			&  &  &  &  &\multicolumn{2}{c}{LTE} & \multicolumn{2}{c}{non-LTE}  \\ 			
			Star & \teff & log~g & [Fe/H] & \vt & logA$_{\rm Ti~I}$ & logA$_{\rm Ti~II}$ & logA$_{\rm Ti~I}$ & logA$_{\rm Ti~II}$  \\ 	
			\hline\noalign{\smallskip}
			Sun & 5777 & 4.44 & 0.0 & 0.9 &--7.12$\pm$0.06 & --7.05$\pm$0.05 & --7.09$\pm$0.08 & --7.06$\pm$0.05 \\
			HD~84937 & 6350 & 4.09 & --2.18 & 1.7 & --8.81$\pm$0.03 & --8.86$\pm$0.10 & --8.65$\pm$0.03 &--8.82$\pm$0.09 \\	
			HD~140283 & 5780 & 3.70 & --2.43 & 1.3 & --9.34$\pm$0.08 & --9.31$\pm$0.07 & --9.16$\pm$0.08 & --9.27$\pm$0.06\\		
			HD~122563 & 4600 & 1.40 & --2.55 & 1.6 & --9.81$\pm$0.06 & --9.42$\pm$0.08 & --9.60$\pm$0.11 & --9.40$\pm$0.07 \\		
			CD--38~245 & 4850 & 1.80 & --3.70 & 1.7 & --10.80$\pm$0.04& --10.86$\pm$0.11 &--10.43$\pm$0.05 &--10.68$\pm$0.10 \\					
			\hline
		\end{tabular}
	\end{center}
\end{table*} 

\section{Method of spectra calculation}\label{method}

We determined the titanium abundance
from Ti~I and Ti~II lines in the non-LTE, where the population of each level in the model
atom is calculated by  solving the
system of the statistical equilibrium equations coupled with the radiative
transfer equation. The  equations
 are solved in a given model atmosphere with the DETAIL
code developed by Butler and Giddings (1985).
The continuum opacity calculations were updated,
as described by Mashonkina et al. (2011). The level
populations obtained in DETAIL were then used to
compute the line profiles with the synthV\_NLTE
code (Tsymbal et al. 2019). To compare the theoretical
spectrum with the observed one, we use
O. Kochukhov's binmag\footnote{http://www.astro.uu.se/$\sim$oleg/download.html} code. We use the classical
plane-parallel model atmospheres interpolated
from the MARCS grid (Gustafsson et al. 2008).

This  study  applies  the  Ti~I-II model atom from Sitnova et al. (2016) with the modifications concerning computations of rates of hydrogenic collisions.
Sitnova et al. (2016) applied the Drawin formula
with the scaling coefficient \kH\ = 1. In this paper, we
calculated the rate coefficients for the inelastic collision
processes using the approximate, but physically
justified method proposed by Belyaev and Yakovleva
(2017a, 2017b) and based on quantum-mechanical
calculations (see Section~\ref{rates_ti}).
The list of Ti~I and Ti~II lines in the visible range
was adopted from our previous paper (Sitnova 2016).
The atomic data for the transitions, namely, the wavelengths
$\lambda$, oscillator strengths (log gf), and
excitation energies of the lower level (\eexc), were taken from the
VALD database (Kupka et al. 1999; Ryabchikova
et al. 2015). Note that the gf values for Ti~I (Lawler
et al. 2013) and Ti~II (Wood et al. 2013) were obtained
in the same laboratory. Table~\ref{linelist} gives the atomic
data for the Ti~II lines in the UV spectral range
that have been used for the first time to determine
the non-LTE titanium abundances in HD~84937 and
HD~140283. The oscillator strengths were taken
from Wood et al. (2013), Pickering et al. (2001), or
R. Kurucz’s database (kurucz.harvard.edu); for each
line the source is specified in Table~\ref{linelist}.

\begin{table*}[htbp]
	\caption{List of Ti~II lines in the UV range}
	\label{linelist}
	\begin{center}
		\begin{tabular}{rrrcrcc}
			\hline\noalign{\smallskip}
			\multicolumn{1}{c}{$\lambda$ }  & $E_{\rm exc}$    &  log gf & Ref. & {\small log$\gamma_{r}$} & {\small log$\gamma_{4}/N_e$} & {\small log$\gamma_{6}/N_H$ } \\
			\multicolumn{1}{c}{	\AA } & eV & &  & s$^{-1}$ & s$^{-1}$cm$^{-3}$ & s$^{-1}$cm$^{-3}$ \\
			\hline
			2041.47 & 0.57 & --1.12 & P & 8.41 & --6.38 & --7.84  \\ 
			2043.23 & 0.57 & --1.82 & K & 8.41 & --6.38 & --7.84  \\ 
			2054.53 & 0.61 & --0.87 & P & 8.41 & --6.38 & --7.84  \\ 
			2135.71 & 1.18 & --1.56 & P & 8.42 & --6.35 & --7.83  \\ 
			2162.68 & 1.24 & --0.49 & P & 8.42 & --6.35 & --7.83  \\ 
			2229.24 & 1.08 & --1.21 & P & 8.41 & --6.38 & --7.84  \\ 
			2261.19 & 1.89 &  0.21 & K & 8.52 & --6.34 & --7.83  \\ 
			2534.62 & 0.12 & --0.93 & W & 8.35 & --6.47 & --7.86  \\ 
			2884.10 & 1.13 & --0.23 & W & 8.45 & --6.54 & --7.83  \\ 
			3046.68 & 1.16 & --0.81 & W & 8.32 & --6.44 & --7.86  \\ 
			3148.04 & 0.00 & --1.22 & P & 8.16 & --6.41 & --7.85  \\ 			
			\hline
		\end{tabular}
	\end{center}
	P -- Pickering et al. (2001), K -- Kurucz database (kurucz.harvard.edu), W -- Wood et al. (2013).\\
\end{table*}  %

\section{Statistical equilibrium of Ti~I-II}\label{seti}

In stellar atmospheres with \teff~>~4000~K, titanium
is highly ionized. For example, $\rm{N_{Ti~II}/N_{Ti~I}} \simeq 10^2$
 everywhere in the solar atmosphere. Because
of the low Ti~I number density compared to Ti~II,
a slight deviation of the mean intensity of ionizing
radiation from the Planck function results in a noticeable
deviation of the Ti~I number density from
the equilibrium one. This mechanism of departures
from LTE is called overionization and leads to an
underpopulation of atomic levels compared to LTE.
For Ti~II, as for the majority species, the
departures from LTE are small and ruled by
bound-bound transitions.

\begin{figure*}  
	\resizebox{80mm}{!}{\includegraphics[trim=1.20cm 0.00cm 0.9cm 0.0cm,clip]{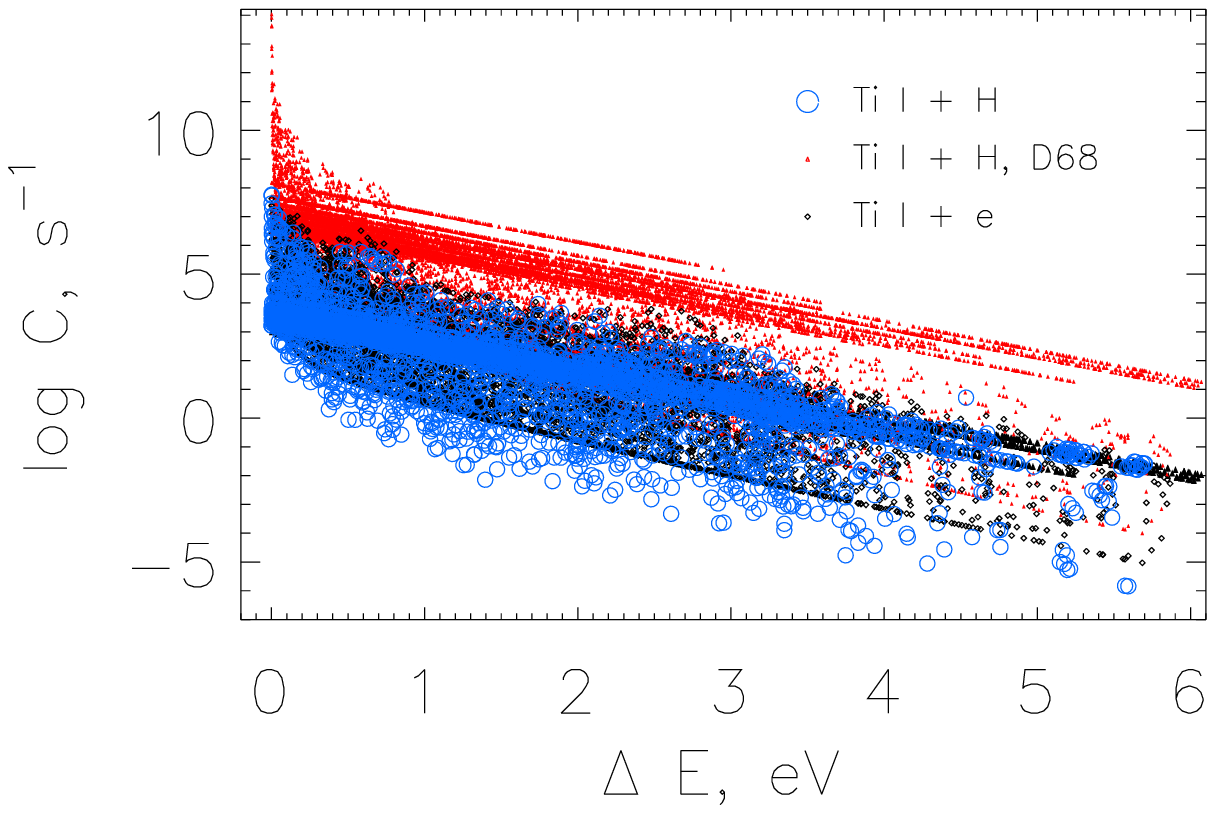}}
	\resizebox{80mm}{!}{\includegraphics[trim=1.77cm 0.00cm 0.0cm 0.0cm,clip]{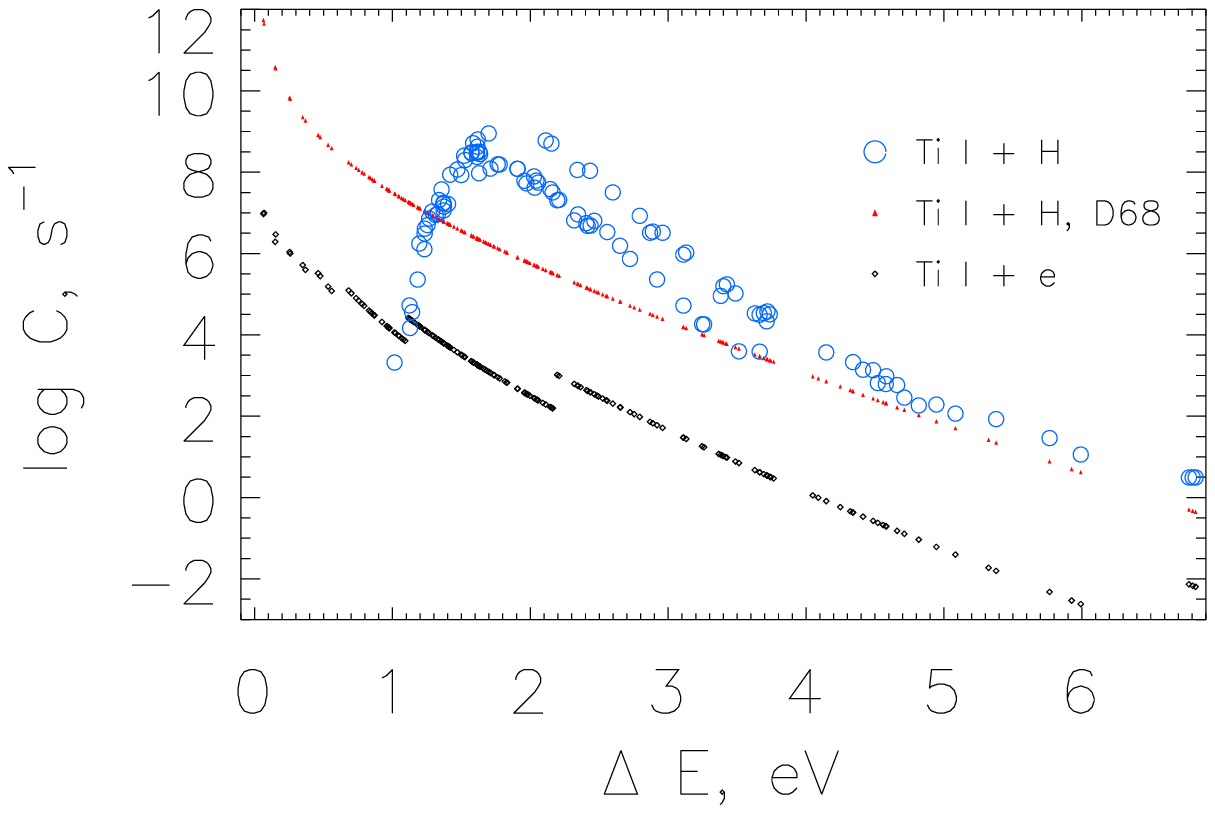}}
	\resizebox{80mm}{!}{\includegraphics[trim=1.20cm 0.00cm 1.2cm 0.0cm,clip]{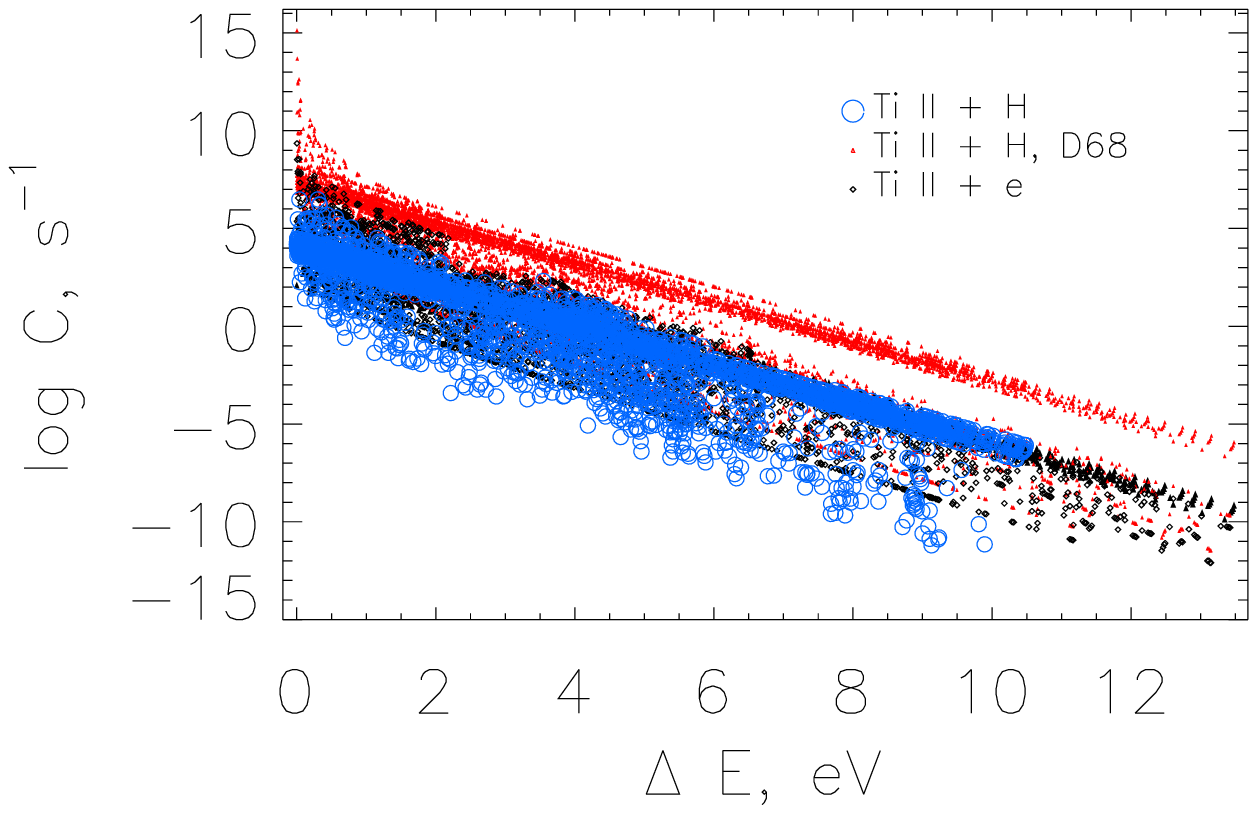}}
	\resizebox{80mm}{!}{\includegraphics[trim=1.20cm 0.00cm 1.2cm 0.0cm,clip]{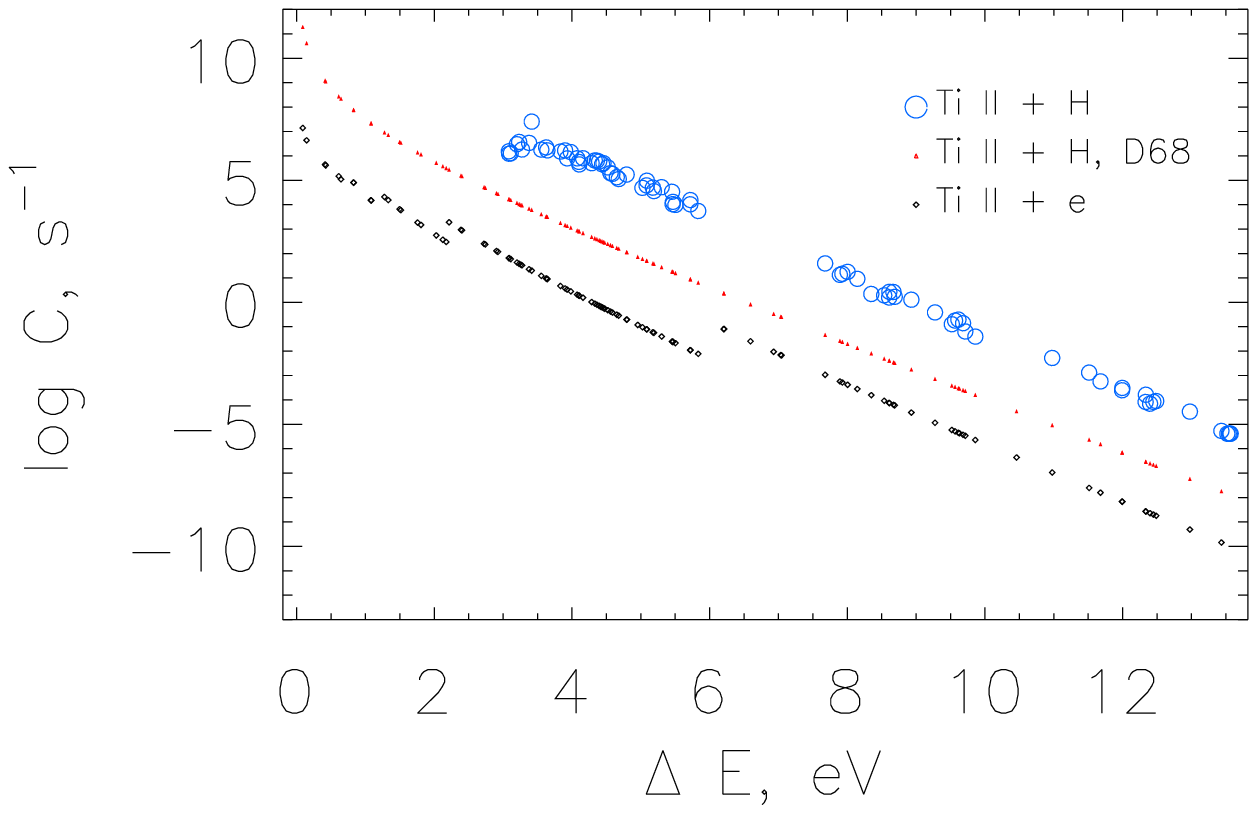}}
	
	\caption{
		Left column: the excitation rates for the Ti~I (upper row) and Ti~II (lower row) transitions in inelastic
		collisions with hydrogen atoms (circles) and electrons (black diamonds). The excitation rates in collisions with hydrogen atoms
		calculated from the Drawin and Takeda formulas (red triangles) are shown for comparison. Right column: the rates of ion-pair
		formation in collisions with hydrogen atoms and electron-impact ionization using analogous symbols. The data are shown for
		a temperature of 5000 K, an electron number density log$N_e$ = 11.9 and a number density of hydrogen atoms log$N_H$ = 16.8
		 } 
	\label{rates}
\end{figure*}

Figure~\ref{rates} shows the excitation and ion pair formation rates in collisions with  H  atoms together with electronic rates, under physical conditions corresponding to the line formation depth in the atmospheres of cool metal-poor stars. 
For comparison, we show the Drawinian rates for inelastic collisions with hydrogen atoms, which were applied in earlier studies due to the absence of accurate data. 
We used the Drawin (1968,
1969) formula to calculate the rates of radiatively
allowed transitions and the approach proposed by
Takeda (1994) for radiatively forbidden transitions.
This approach implies that the rate of inelastic collisions
with hydrogen atoms (C$_{\rm H}$) is calculated via
the electron collision rate (C$_{\rm e}$) from the relation 
C$_{\rm H}$ = C$_{\rm e}\sqrt{m_{\rm e}/m_{\rm H}}N_{\rm H}/N_{\rm e}$, where m$_{\rm H}$, m$_{\rm e}$ and $N_{\rm H}$, $N_{\rm e}$ are
the masses and number densities of hydrogen atoms
and electrons, respectively. For Ti~I and Ti~II, the
excitation rates in collisions with hydrogen atoms
derived in this paper are equal in order of magnitude
to those in collisions with electrons and are lower
than the Drawinian rates by three orders of magnitude.
The ion-pair formation rates in collisions with hydrogen
atoms exceed the electron-impact ionization
rates approximately by three orders of magnitude at
ionization energies greater than 2 eV.

The rates obtained in this paper differ fundamentally
from those that we applied previously, since the former are based on a physically realistic assumption.
First, the process of conversion, for example, of Ti~I
into Ti~II, occurs via charge exchange, 
Ti~I + H $\leftrightarrow$ Ti~II + H$^-$ , while the Drawin formula was used
for ionization,  Ti~I + H $\leftrightarrow$ Ti~II + H + e. Second,
when the excitation rates are calculated from the
Drawin and Takeda formulas, all levels in the model
atom are coupled  by collisions, while
quantum-mechanical calculations predict that the
transitions in inelastic collisions with hydrogen atoms
occur between the levels that satisfy the selection
rules for a molecular symmetry. 
A decrease in the number of transitions, for which the inelastic collisions with hydrogen atoms are efficient, leads to an increase of the departures from LTE.

\begin{figure*}  
	\resizebox{80mm}{!}{\includegraphics[trim=0.90cm 0.60cm 1.3cm 1.4cm,clip]{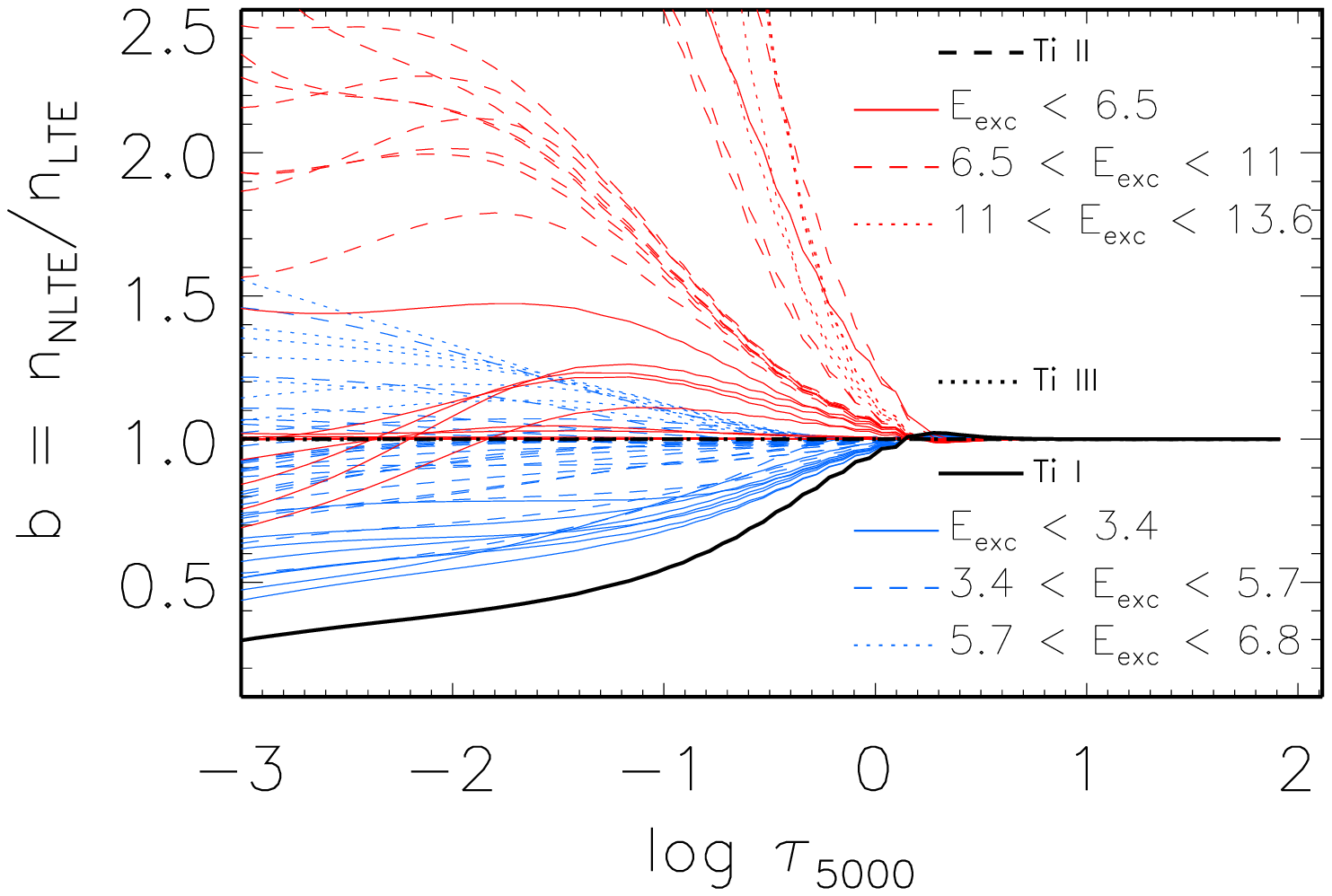}}
	\resizebox{80mm}{!}{\includegraphics[trim=0.90cm 0.60cm 1.3cm 1.4cm,clip]{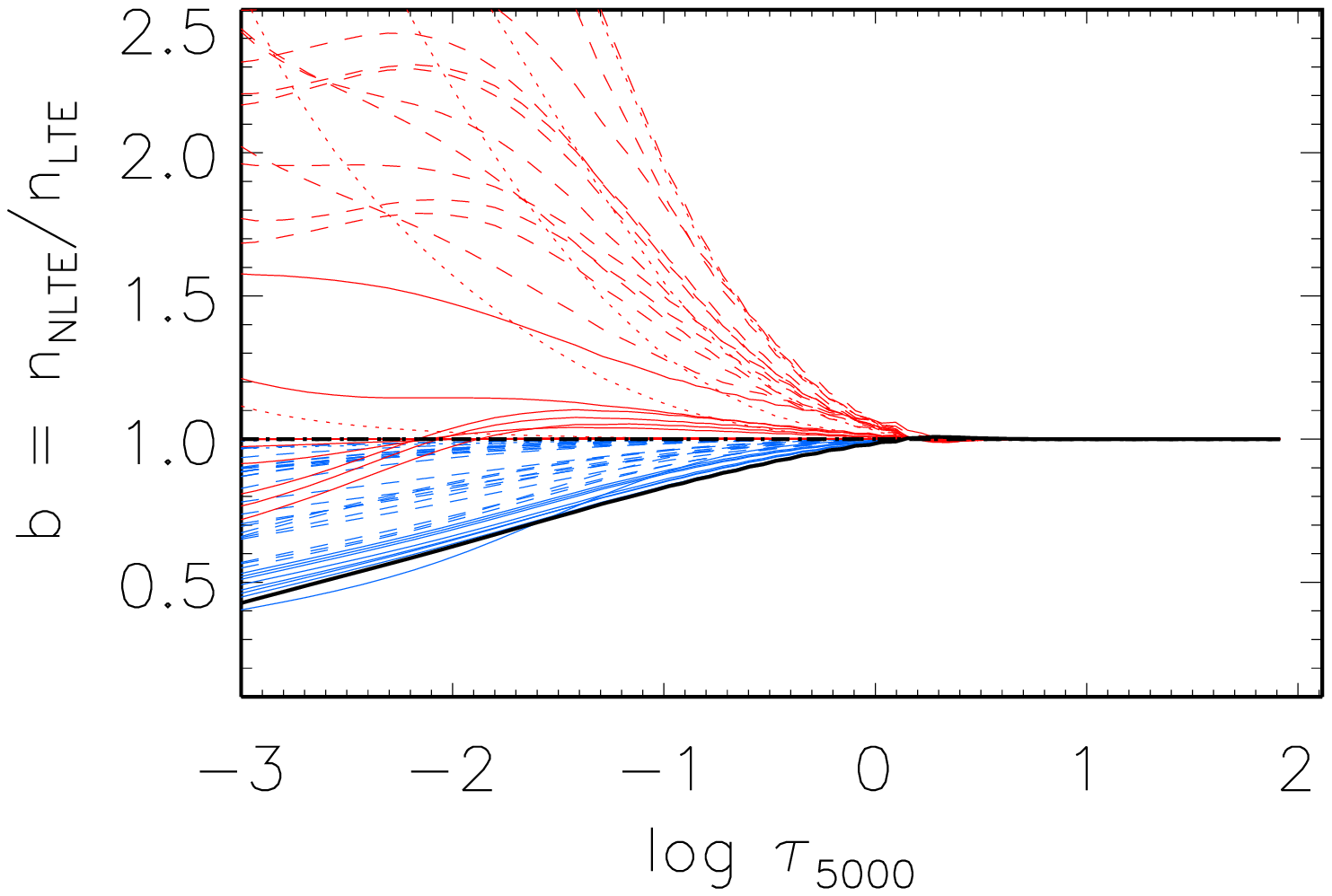}}%
	\caption{		The departure coefficients for Ti~I and Ti~II levels in the model atmosphere with \teff\ = 6350, log g = 4.09, and [Fe/H] = --2.1
		calculated with the rates of inelastic collisions with hydrogen atoms from this paper (left panel) and from approximate formulas (right panel).
		Different lines indicate the departure coefficients of the levels in different ranges of excitation energies.
		} 
	\label{bf}
\end{figure*}

The deviation of the level populations from the
equilibrium ones are characterized by the departure coefficients,
b$_i$ = \nnlte/\nlte, where \nnlte\ and \nlte\, are the
populations of level i in the non-LTE and LTE cases,
respectively. Figure~\ref{bf} shows the departure coefficients for Ti~I
and Ti~II levels in the model atmosphere with \teff\ = 6350~K, log~g = 4.09, [Fe/H] = --2.1 calculated with
the rates of inelastic collisions with hydrogen atoms
from this paper and from the approximate formulas. The
Ti~I levels with an excitation energy \eexc\ < 5 eV are
depleted as a result of overionization by UV radiation.
The lower the excitation energy of the level, the
greater its underpopulation, which leads to a weakening
of the Ti~I lines due to a decrease in the line opacity
($\kappa_{\nu} \simeq b_l$, where b$_l$  is the departure coefficient of the lower level).
The Ti~II levels, on the contrary, are overpopulated
due to radiative pumping, the more so, the greater
their excitation energy. On the one hand, the line
opacity increases due to the lower-level overpopulation,
but, on the other hand, the source function exceeds
the Planck function at the transition frequency
due to a greater overpopulation of the upper level
compared to the lower one (S$_{\nu}$/B$_{\nu} \simeq b_u/b_l$, where S$_{\nu}$
is the source function, B$_{\nu}$ is the Planck function, $b_u$ and $b_l$
 are the departure coefficients of the upper and lower levels,
respectively). Therefore, for Ti~II lines, non-LTE can
lead to both line weakening and strengthening. The
ground level of the majority species, Ti~II,
retains its equilibrium population in all atmospheric
layers.

With the application of new data, the departures
from LTE are strengthened due to a decrease in the
collision rates. Besides with the increase of the
departures from LTE, it is important that the coupling
between levels with close energies weakens. This
effect is most pronounced for the departure coefficient of the ground  level of Ti~I,
which became separated from the departure coefficient of the levels with higher excitation
energy (Fig.~\ref{bf}).

\section{Titanium abundance in the sample stars} \label{abundances}

For the Sun and four metal-poor stars, we determined
the non-LTE and LTE titanium abundances
from Ti~I and Ti~II lines. The application of accurate
data led to an increase in the mean non-LTE abundance
from Ti~I and Ti~II lines for all sample
stars. For the Sun, the mean non-LTE abundance
increased only by 0.01 dex for both Ti~I and Ti~II. Despite the small change in the mean non-LTE abundance, for individual Ti~I lines the difference
in collision rates leads to changes in the non-LTE
abundance up to 0.09 dex in absolute value. In the Sun, the non-LTE abundance
corrections 
(\dnlte\ = \abnlte\ -- \ablte) 
for Ti~I and Ti~II
lines do not exceed 0.1 dex in absolute value.

\begin{figure}  
	\resizebox{80mm}{!}{\includegraphics[trim=0.0cm 1.3cm 0.0cm 0cm,clip]{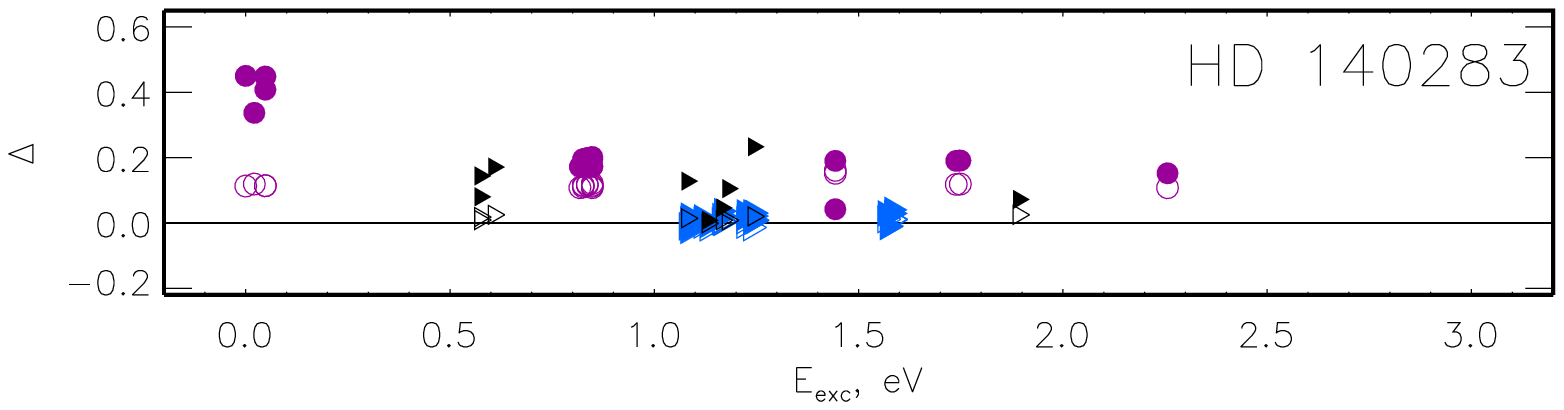}}
	\resizebox{80mm}{!}{\includegraphics[trim=0.0cm 1.3cm 0.0cm 0.7cm,clip]{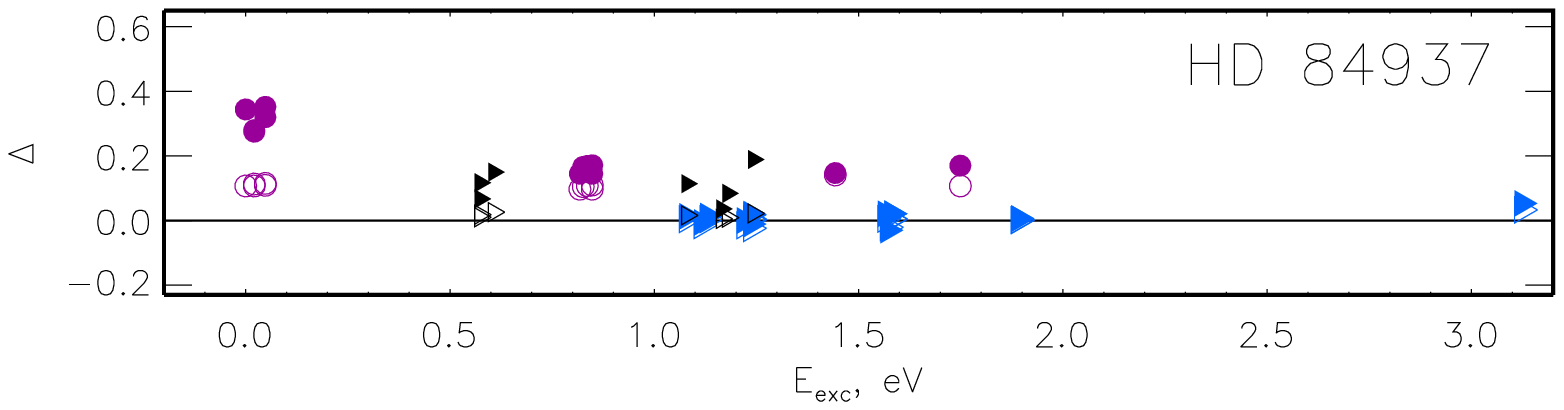}}
	\resizebox{80mm}{!}{\includegraphics[trim=0cm 1.3cm 0cm 0.70cm,clip]{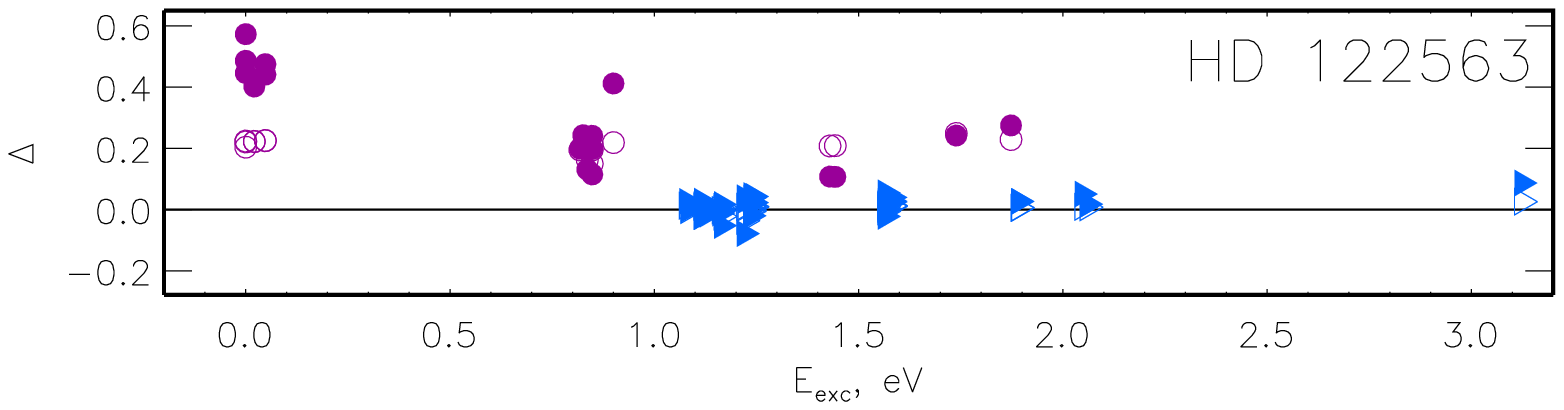}}
	\resizebox{80mm}{!}{\includegraphics[trim=0cm 0.0cm 0cm 0.70cm,clip]{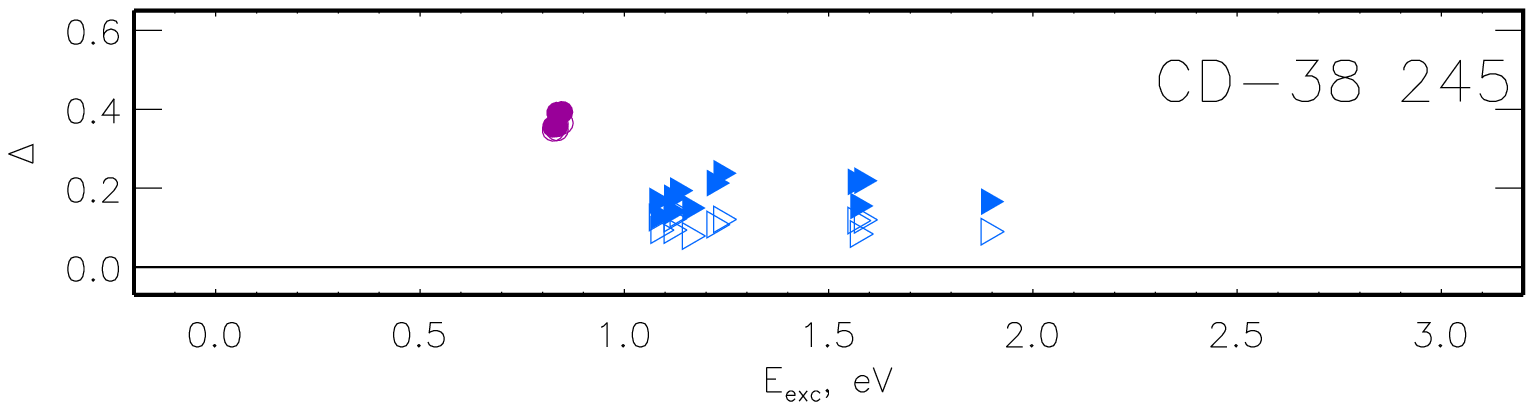}}
	\caption{The non-LTE abundance corrections for different Ti~I (circles) and Ti~II (triangles) lines in our metal-poor
		stars versus lower-level excitation energy. The corrections derived with the accurate rates of inelastic collisions with hydrogen
		atoms and those calculated from the approximate Drawin and Takeda formulas are indicated by the filled and open symbols,
		respectively. The non-LTE corrections for Ti~II lines in the UV spectral range are indicated by smaller triangles.} 
	\label{delta}
\end{figure}

For our metal-poor stars,  the non-LTE corrections for different lines calculated with the
accurate and approximate rates are shown in Fig.~\ref{delta}.
In the non-LTE calculations with
the approximate rates, the non-LTE corrections for the lines forming in
transitions with different lower-level excitation energies are similar.
For most lines, the application of accurate
data led to an increase in the non-LTE corrections.
We obtained the largest change in the non-LTE
abundance for the Ti~I lines forming in the transitions
from the ground level. For these lines, the non-LTE
corrections increase by up to 0.3 dex in HD~140283
and HD~122563. At the same time, with the new data
the non-LTE corrections themselves in different stars
are 0.3--0.5 and 0.1--0.2 dex for the lines from the
ground level and levels with \eexc\ > 0.8 eV, respectively.

For Ti~II lines in the visible spectral range, the
changes in non-LTE abundance  do
not exceed 0.1 dex, while for Ti~II lines in the UV
range we obtained significant changes, for example,
0.2 dex for Ti~II 2162 \AA\ in HD~140283. The non-LTE corrections themselves are either positive, up to
0.25 dex, or small negative, no more than 0.1 dex
in absolute value for strong lines with an equivalent
width 100 m\AA\ < EW < 120 m\AA.

\begin{figure}  
	\resizebox{80mm}{!}{\includegraphics{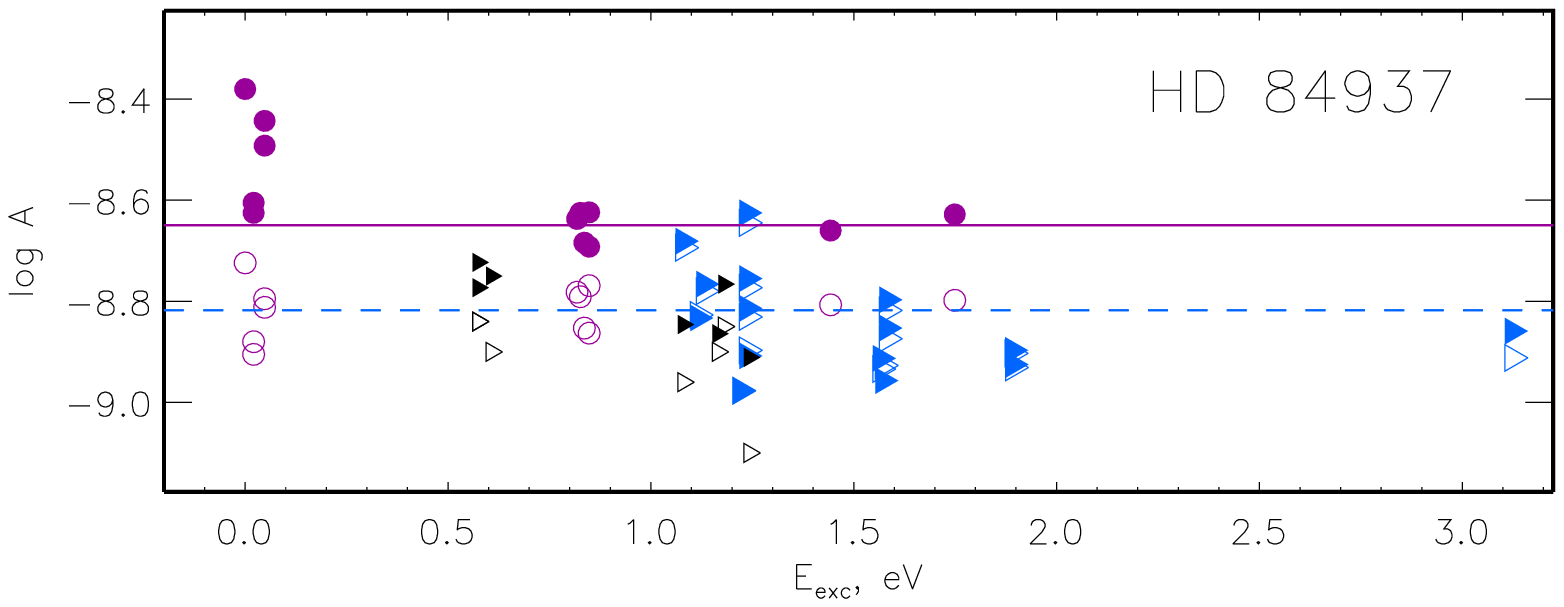}}		
	\resizebox{80mm}{!}{\includegraphics{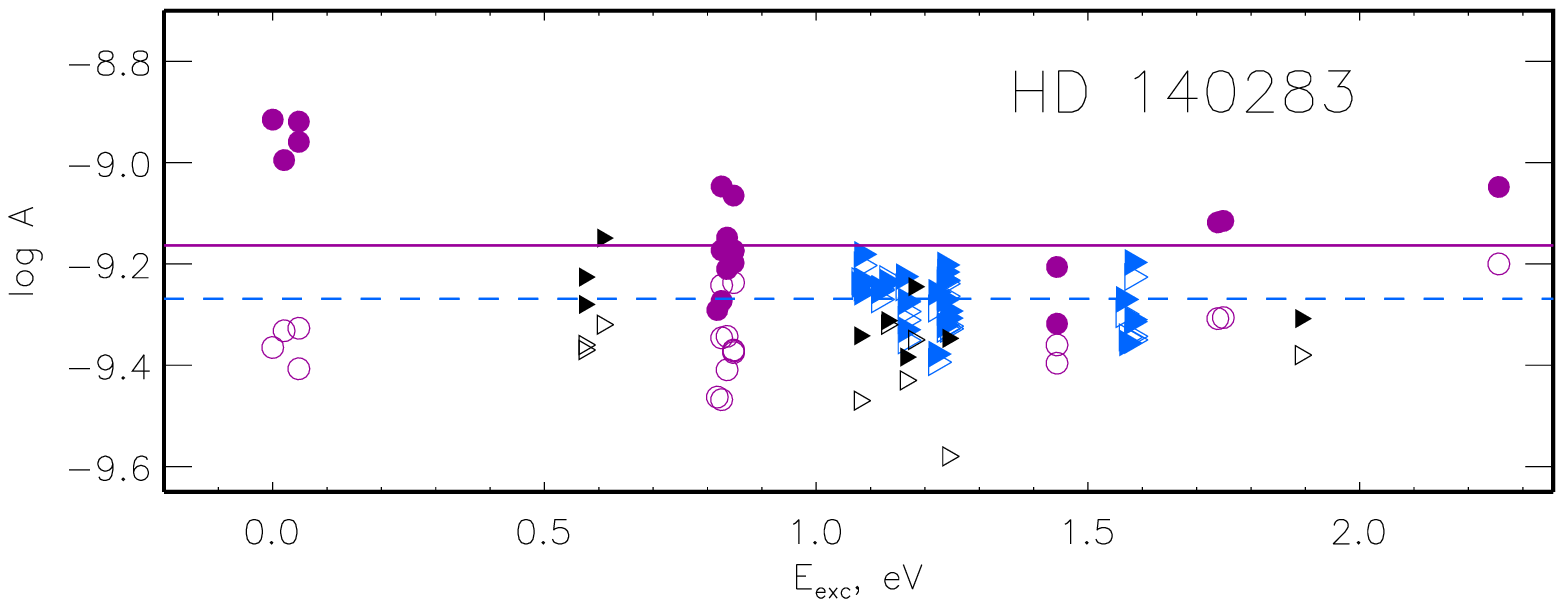}}
	\resizebox{80mm}{!}{\includegraphics{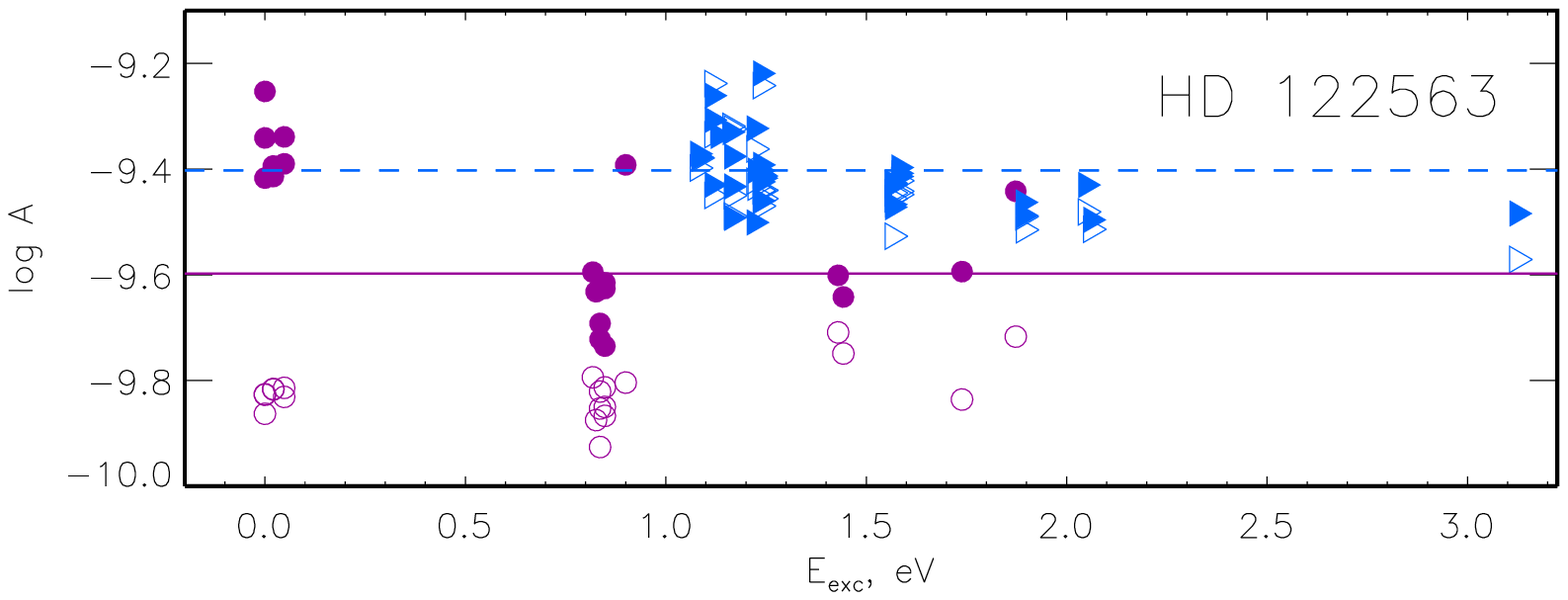}}
	\resizebox{80mm}{!}{\includegraphics{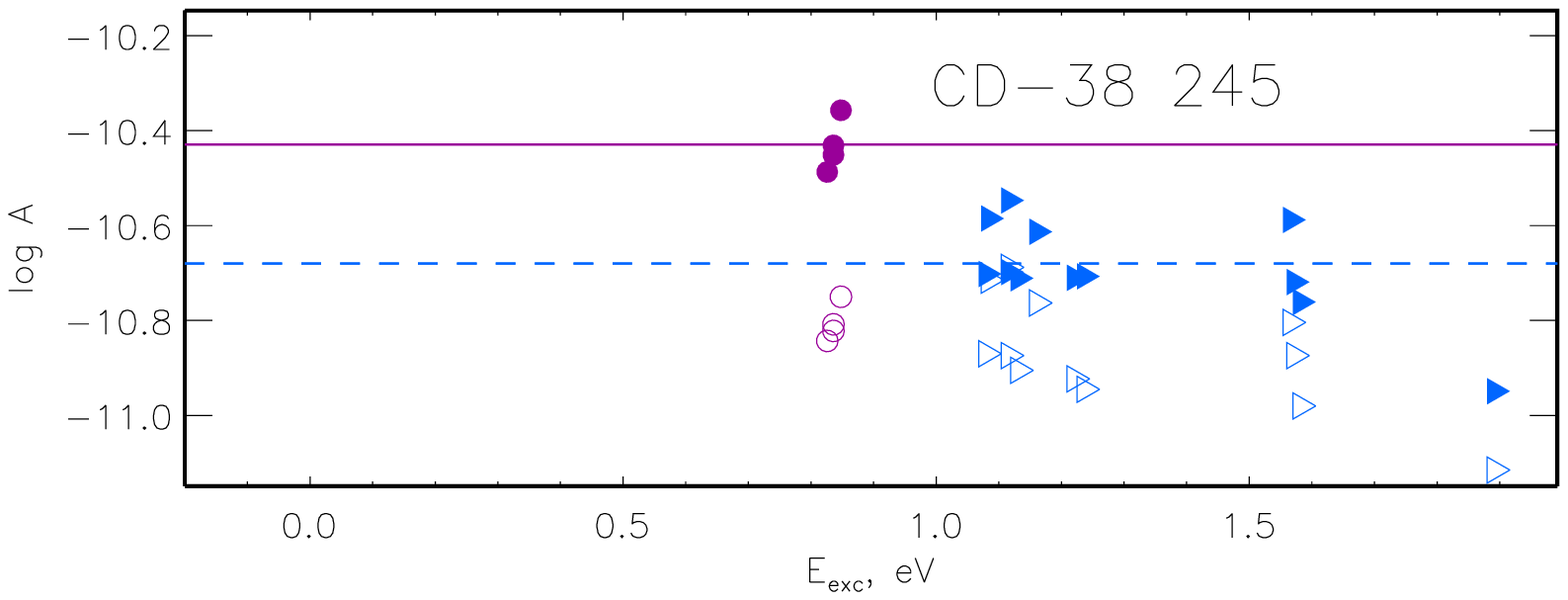}}		
	\caption{
		Titanium abundances in our metal-poor stars from different Ti~I (circles) and Ti~II (triangles) lines versus
		lower-level excitation energy. The non-LTE and LTE abundances are indicated by the filled and open symbols, respectively.
		For HD 84937 and HD 140283 the titanium abundances from Ti~II lines in the UV spectral range are indicated by smaller
		triangles. The mean non-LTE abundance from Ti~I (without  lines originating from the ground state) and Ti~II lines is indicated by the solid
		and dashed lines, respectively.
		 } 
	\label{abund}
\end{figure}

Figure~\ref{abund} shows the derived non-LTE and LTE
abundances from individual lines in the metal-poor
sample stars. The changes in the rates led to a
discrepancy in the abundances from Ti~I lines from
the ground level and with an excitation energy \eexc\ >
0.8 eV in all four stars. Such a behavior is observed for
Fe~I lines with \eexc\ < 1.5 eV and can be interpreted as
a signature of convection (3D-effects, see, e.g.,
Collet et al. 2007; Dobrovolskas et al. 2013; Amarsi
et al. 2019). These lines are not recommended 
for abundance determinations with classical model
atmospheres. In this paper, the mean abundance
from Ti~I lines was calculated neglecting  the
lines forming in the transitions from the ground level.
Table~\ref{param_tab} gives the non-LTE and LTE abundances from
Ti~I and Ti~II lines. For HD~140284 and HD~84937,
for which UV spectra are available, we found the non-LTE
abundances from Ti~II lines in the visible and UV
ranges to be consistent.

For the sample stars, we compare the titanium
abundances calculated separately from Ti~I and Ti~II
lines. For the Sun, non-LTE leads to an agreement
between the abundances from lines of different ionization
stages. The mean abundance difference between
Ti~I and Ti~II is $\Delta_{\rm Ti~I - Ti~II}$ = --0.07~dex$\pm$0.08 and
--0.03~dex$\pm$0.09 in LTE and non-LTE, respectively.
For HD~140283, HD~84937, and CD-38~245, the
abundances from two ionization stages agree in
LTE. For HD~122563,  the LTE difference is
$\Delta_{\rm Ti~I -Ti~II}$ = --0.4~dex. In non-LTE, the abundances
from Ti~I and Ti~II lines do not agree and $\Delta_{\rm Ti~I - Ti~II}$ = 0.10~dex$\pm$0.10, 0.17~dex$\pm$0.09, --0.20~dex$\pm$0.13, and 0.25~dex$\pm$0.11 for HD~140283, HD~84937,
HD~122563, and CD-38~245, respectively. Such a
behavior cannot be explained by the errors in
the atmospheric parameters. For example,
for CD-38~245, where the uncertainties in \teff\ and
log g are largest compared to the other sample stars, a
decrease in \teff\ by 250 K or an increase in log g by
0.3 allows the discrepancy to be reduced only by 0.15
and 0.09 dex, respectively. However, such changes in
parameters would lead to an implausible evolutionary
status of CD-38~245 that does not correspond to its
metallicity, age, and position on the corresponding
isochrone.

The problem of a discrepancy in the non-LTE
abundances from Ti~I and Ti~II lines in metal-poor
stars was discussed in the literature (Bergemann
2011; Sitnova et al. 2016; Sitnova 2016).
Table~\ref{deltaion} gives the abundance difference $\Delta_{\rm Ti~I -Ti~II}$ for
three sample stars obtained in this paper, our previous
paper (Sitnova et al. 2016), and Bergemann (2011).
For comparison, we also provide the analogous
quantities for iron $\Delta_{\rm Fe I - Fe II}$ derived by Mashonkina
et al. (2019) with the  accurate
quantum-mechanical data for Fe~I + H and Fe~II + H
collisions. For titanium, no agreement between the
non-LTE abundances from different ionization stages
was obtained in any of the papers. Bergemann (2011)
suggested that the problem could lie in an insufficient
accuracy of the atomic data. Since then, a significant
progress has been achieved in the atomic data for
Ti~I and Ti~II line calculations, and (i) laboratory measurements
of the oscillator strengths for Ti~I (Lawler
et al. 2013) and Ti~II (Wood et al. 2013) transitions,
(ii) accurate quantum-mechanical calculations of the
photoionization cross sections for Ti~I (Nahar 2015)
and Ti~II (K. Butler, private communication 2015),
and (iii) accurate quantum-mechanical calculations of
the transition rate coefficients in inelastic collisions
with hydrogen atoms (this paper) have appeared.
However, as can be clearly seen from Table~\ref{deltaion}, the
application of these data has not solved the problem
of a discrepancy in the abundances from Ti~I and
Ti~II lines in metal-poor stars. Note
that there is also a hint of a higher abundance from
Fe~I lines than that from Fe~II in non-LTE for iron
(Mashonkina et al. 2011, 2019). However, this effect
for iron is weaker than for titanium, possibly, due to
smaller departures from LTE.

The problem refers more to the Ti~I lines than to
Ti~II. The elemental abundance ratios in Milky Way
halo stars argue for this conclusion. From a non-LTE
analysis of Mg~I, Ca~I, Ti~II, and Fe~II lines Zhao
et al. (2016) found that for stars with [Fe/H] < --1
the elemental abundance ratios [Mg, Ca, Ti/Fe] $\simeq$
0.3. For HD~84937, HD~140283, and HD~122563
[Ti~II/Fe] = 0.44, 0.25, and 0.26, respectively, and are
close to  [$\alpha$/Fe] from Zhao et al. (2016), while from
Ti~I with [Ti~I/Fe] = 0.61, 0.36, and 0.06 the elemental
abundance ratios shows larger deviations from the mean Galactic [$\alpha$/Fe].
As before, we conclude that Ti~II lines in metal-poor stars
give more reliable non-LTE abundance than that
obtained from Ti~I lines.
Depending on the stellar effective temperature, the
abundance from Ti~I lines is either overestimated, as
in the case of HD~84937 with \teff\ = 6350 K and, to a
lesser extent, HD~140283 with \teff\ = 5780 K, or, on
the contrary, underestimated, as in the case of the
giant HD~122563 with \teff\ = 4600 K. It should be noted
that the titanium abundance in CD-38~245 does not
follow this trend. The metallicity of this star is lower
than that for the other three metal-poor stars by more
than an order of magnitude. More data for stars in
different metallicity ranges are needed to understand
the  behavior of the non-LTE abundance
discrepancy as a function of atmospheric parameters.

\begin{table*}[htbp]
	\caption{Comparison of the non-LTE abundances from lines of different ionization stages in the sample stars}
	\label{deltaion}
	\tabcolsep1.50mm
	\begin{center}
		\begin{tabular}{lrrrccc}
			\hline\noalign{\smallskip}	
			& \multicolumn{3}{c}{$\Delta_{\rm Ti~I -Ti~II}$} & $\Delta_{\rm Fe I - Fe II}$ & [Ti~I/Fe] & [Ti~II/Fe]\\
			Star  &	S20 & S16 & B11 &  M19 & \multicolumn{2}{c}{S20 + M19} \\ 	
			\hline\noalign{\smallskip}	
			HD~84937 & 0.17  & 0.15  & 0.24  & 0.10 & 0.61 & 0.44 \\	
			HD~140283 & 0.10  & 0.09  & 0.16  & 0.09 & 0.36 & 0.25  \\
			HD~122563 &--0.20  &--0.18  & --0.10  & --0.07  & 0.06 & 0.26 \\		
			\hline
			\multicolumn{7}{c}{S20 -- this paper, S16 -- Sitnova et al. (2016),  } \\
			\multicolumn{7}{c}{B11 -- Bergemann (2011), M19 -- Mashonkina et al. (2019).} \\
		\end{tabular}
	\end{center}
\end{table*} 

\section{Conclusions}

In this paper, we made an attempt to solve the
problem of a discrepancy in the non-LTE abundances
from Ti~I and Ti~II lines in metal-poor stars by applying
 accurate data to for the inelastic
collisions with hydrogen atoms. For this purpose,
we calculated the rate coefficients for bound-bound
transitions in inelastic collisions of titanium atoms
and ions with hydrogen atoms and for the following
charge-exchange processes: Ti~I + H $\leftrightarrow$ Ti~II + H$^-$ and Ti~II + H $\leftrightarrow$ Ti~III + H$^-$. The influence of accurate
data on non-LTE abundance determinations was
tested for the Sun and four metal-poor stars.

The application of the derived rate coefficients led
to an increase in the departures from LTE and an
increase in the titanium abundance compared to what
is obtained using approximate formulas. For the
Sun, the mean non-LTE abundance increased only by
0.01 dex from both Ti~I and Ti~II lines. In metal-poor stars, we found a
significant change in the non-LTE abundance, up to
0.3 dex, for the Ti~I lines forming in the transitions
from the ground level. This led
to a discrepancy in the abundances from Ti~I lines
from the ground level and with an excitation energy
\eexc\ > 0.8 eV in the  sample stars.

For our metal-poor stars, the mean non-LTE
abundance from Ti~II lines increased by a few
hundredths, from 0.01 to 0.07 dex, depending on the
atmospheric parameters. For Ti~II lines in the UV
range, the non-LTE corrections are significant and
can reach 0.2 dex in dwarfs with [Fe/H] = --2.
In HD~84937 and HD~140283, we found consistent non-LTE
abundances   from
Ti~II lines in the visible and UV ranges.

For the Sun, non-LTE leads to agreement between
the abundances from Ti~I and Ti~II lines. The
mean abundance difference between Ti~I and Ti~II is
$\Delta_{\rm Ti~I -Ti~II}$ = $-0.07$~dex and $-0.03$~dex in LTE and non-LTE, respectively. For our metal-poor stars, there is
a discrepancy in the non-LTE abundances from Ti~I
and Ti~II lines; $\Delta_{\rm Ti~I -Ti~II}$ = 0.10~dex, 0.17~dex, --0.20~dex, and
0.25 dex for HD~140283, HD~84937, HD~122563, and
CD-38~245, respectively. The problem of a discrepancy
in the abundances from Ti~I and Ti~II lines in
metal-poor stars known in the literature, apparently,
cannot be solved via an improvement of
the rates of  inelastic processes in collisions with hydrogen atoms
in non-LTE calculations with classical model atmospheres.

{\bf Acknowledgments:}
S.A. Yakovleva and A.K. Belyaev gratefully acknowledge support from the Ministry
of Science and Higher Education of the Russian
Federation (project
nos. 3.5042.2017/6.7 and 3.1738.2017/4.6). We
are grateful to K. Fuhrmann for the spectra and
O. Kochukhov for the binmag code. We used the
VALD, MARCS, and ASTRAL databases.


%

\end{document}